\documentclass[aps,prl,onecolumn,showpacs,amsmath,amssymb,floatfix,footinbib,superscriptaddress]{revtex4-1}
\usepackage{amsmath,natbib,graphicx,subfigure,amssymb,graphics,amsmath,mathrsfs,CJK,color}
\usepackage{multirow,fancyhdr,color,bm,tabularx,psfrag,dcolumn}
\usepackage[dvipdfm,colorlinks=true,linkcolor=blue,urlcolor=blue,citecolor=blue]{hyperref}
\usepackage{tikz,xcolor,hyperref}
\definecolor{lime}{HTML}{A6CE39}
\DeclareRobustCommand{\orcidicon}{%
    \begin{tikzpicture}
    \draw[lime, fill=lime] (0,0)
    circle [radius=0.16]
    node[white] {{\fontfamily{qag}\selectfont \tiny ID}};\draw[white, fill=white] (-0.0625,0.095)
    circle [radius=0.007];
    \end{tikzpicture}
    \hspace{-2mm}}
\foreach \x in {A, ..., Z}
{\expandafter\xdef\csname orcid\x\endcsname{\noexpand\href{https://orcid.org/\csname orcidauthor\x\endcsname}{\noexpand\orcidicon}}}

\usepackage[left]{lineno}
\begin{document}

\title{Negative refraction without absorption via both coherent and incoherent fields in a four-level left-handed atomic system\footnote{Supported by National
Natural Science Foundation of China (Grant No.60768001 and
No.10464002).}}

\author{Shun-Cai Zhao\orcidA{}}
\email[Corresponding author: ]{zscnum1@126.com.}
\affiliation{School of Materials Science and Engineering, Nanchang University, Nanchang 330031, PR China}
\affiliation{Engineering Research Center for Nanotechnology, Nanchang University, Nanchang 330047, PR China}
\affiliation{Institute of Modern Physics,Nanchang University, Nanchang 330031, PR China}

\author{Zheng-Dong Liu}
\email[Corresponding author: ]{lzdgroup@ncu.edu.cn.}
\affiliation{School of Materials Science and Engineering, Nanchang University, Nanchang 330031, PR China}
\affiliation{Engineering Research Center for Nanotechnology, Nanchang University, Nanchang 330047, PR China}
\affiliation{Institute of Modern Physics,Nanchang University,Nanchang 330031, PR China}

\author{Qi-Xuan Wu}
\affiliation{College English department,Hainan University, Danzhou 571737, PR China}


\begin{abstract}
This paper attempts a probe into negative refraction without
absorption by means of an incoherent pump field and a strong
coherent field coupling the dense four-level atomic system.With the
application of the incoherent pump field to manipulate the
populations in atomic levels and the variable strong coherent field
to create quantum coherence, the constraint condition of two equal
transition frequencies responding to the probe field in the atomic
system isn't required.And these lead to the propagation transparency
and strong magnetic response of the probe field,left-handedness with
vanishing absorption in the atomic system.However,an excessive
coherent field intensity would increase the absorption.
\begin{description}
\item[PACs]{42.50.Gy}
\item[Keywords]{negative refraction; without absorption; left-handed; coherent
field; incoherent field.}
\end{description}
\end{abstract}

\maketitle
\section{Introduction}

 Negative refraction of electromagnetic radiation[1]has recently
attracted considerable attention because of its surprising and
counterintuitive electromagnetical and optical effects,such as the
reversals of both Doppler shift and Cherenkov effect,negative
refraction[2],amplification of evanescent waves[3]and subwavelength
focusing [3-5], negative Goos-H$\ddot{a}$nchen shift [6] and
quenching spontaneous emission[7,8]and so on[9].And material with
negative refraction index intrigues the researchers because of its
many significant potential applications.The`` perfect lens ''is one
of them.Since a slab of such materials has an ability to focus all
frequency components of a two-dimensional image, it may become
possible to make a `` perfect lens ''in which imaging resolution
does not limited by the diffraction limit[3].Up to now, there have
been several approaches to the realization of negative refractive
index materials, including artificial composite metamaterials
[10-11], photonic crystal structures [12],transmission line
simulation[13] and chiral media[14-15] as well as photonic resonant
materials(coherent atomic vapour)[16-19].In such a type of negative
refractive materials, negative refraction is usually accompanied by
a strong absorption especially towards higher frequencies. Thus,the
realization of negative refraction material without absorption is of
great significance.And some effort[20-23]has been made to realize
negative refraction without absorption.K$\ddot{a}$stel et al[20]
realized negative refraction with minimal absorption in a dense
atomic gas via electromagnetically induced chirality. The key
ingredient of the scheme is the electromagnetic chirality that
results from coherently coupling a magnetic dipole transition with
an electric dipole transition via atomic coherence induced by the
two-photon resonant Raman transitions. In Ref.[21],K$\ddot{a}$stel
et al also discussed negative refraction with reduced absorption due
to destructive quantum interference in coherently driven atomic
media.The negative refraction with deeply depressing absorption and
without simultaneously requiring both negative electric permittivity
and magnetic permeability (left-handedness)[2]was obtained by F.L.
Li[23].Ref.[23] shows this at the ideal situation that the two
chirality coefficients have the same amplitude but the opposite
phase.

In this paper we propose an indirect coupling way of the atoms
responding to the probe field via an incoherent pump field and a
coherent coupling field.With the application of the incoherent pump
field to manipulate the population distribution of each level and
the variable coherent field to create quantum coherence,the magnetic
response is amplified and the probe field propagates transparently.
The atomic system displays negative refraction without absorption
and left-handedness.The constraint condition of two equal transition
frequencies responding to the probe field does not require,which is
different from Ref.[18,24].

\section{Theoretical model}

The level configuration of atoms under consideration is shown in
Figure 1. The properties of the four atomic states are as following:
levels $|1\rangle$,$|3\rangle$, and $|4\rangle$ have same parity,
and the parity of level$|2\rangle$ is opposite with theirs. The two
lower levels $|1\rangle$ and $|2\rangle$ have opposite parity and so
$\langle2|$$\hat{\vec{d}}$$|1\rangle$$\neq0$ where $\hat{\vec{d}}$
is the electric dipole operator. The two upper levels,$|3\rangle$
and $|4\rangle$ have the same parity with
$\langle4|$$\hat{\vec{\mu}}$$|3\rangle$$\neq0$ where
$\hat{\vec{\mu}}$ is the magnetic-dipole operator. As observed in
Fig.1,three electromagnetic fields are introduced to couple the four
states:The electric and magnetic components of the probe light
(corresponding Rabi frequency
$\Omega_{p}$$=\frac{\vec{E_{P}}\vec{d_{21}}}{\hbar}$
,$\Omega_{B}$$=\frac{\vec{B_{P}}\vec{\mu_{43}}}{\hbar}$) interact
with the transitions $|2\rangle$ and$|1\rangle$ as well
as$|4\rangle$ and$|3\rangle$, respectively.The incoherent pump field
with pumping rate denoted by $\Gamma$ pumps atoms in level
$|1\rangle$ into upper level $|3\rangle$, and then the atoms decay
into metastable level $|2\rangle$ via rapid nonradiative
transitions, whose decay rate is denoted as $\Gamma_{32}$.The strong
coherent field $\Omega_{c}$ takes two effects on the system:(i)
populations in level $|2\rangle$ are being pumped into level
$|4\rangle$ and causes population reversion gain for magnetic-dipole
transition $|3\rangle$$\leftrightarrow$$|4\rangle$. (ii)quantum
coherence is induced by it and causes energy levels $|2\rangle$ and
$|4\rangle$ split into two dressed sublevels and results in enough
response of the electric part of probe field at certain frequency
extent.

\begin{center}
\begin{figure}[h!]
  \centering
  \includegraphics[width=0.45\columnwidth]{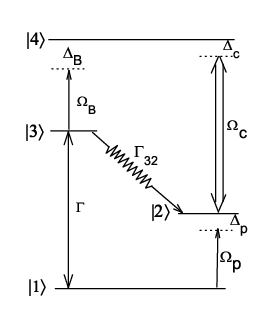}%
  \hspace{0in}%
  \caption{Four-level atomic medium interacting with a coherent
$\Omega_{c}$ , an incoherent pumps $\Gamma$ and a probe field
$\Omega_{p}$.}
\end{figure}\label{Fig.1}
\end{center}

Using the density-matrix approach, the time-evolution of the system
is described as
\begin{equation}
\frac{d\rho}{dt}=-\frac{i}{\hbar}[H,\rho]+\Lambda\rho ,
\end{equation}
Where $\Lambda\rho$ represents the irreversible decay part in the
system.Under the dipole approximation and the rotating wave
approximation the density matrix equations described the system are
written as follows:

\begin{equation}
\dot{\rho_{11}}=\Gamma(\rho_{33}-\rho_{11})+\Gamma_{21}\rho_{22}+\Gamma_{31}\rho_{33}+\Gamma_{41}\rho_{44}+i\Omega_{p}(\rho_{21}-\rho_{12}),
\end{equation}
\begin{equation}
\dot{\rho_{21}}=-(\gamma_{21}+i\Delta_{p})\rho_{21}-i\Omega_{p}(\rho_{22}-\rho_{11})+i\Omega_{c}\rho_{41},
\end{equation}
\begin{equation}
\dot{\rho_{22}}=-\Gamma_{21}\rho_{22}+\Gamma_{32}\rho_{32}+\Gamma_{42}\rho_{44}+i\Omega_{p}(\rho_{12}-\rho_{21})+i\Omega_{c}(\rho_{42}-\rho_{24}),
\end{equation}
\begin{equation}
\dot{\rho_{31}}=-[\gamma_{31}+i(\Delta_{c}+\delta)]\rho_{31}+i\Omega_{p}\rho_{23}+i\Omega_{B}\rho_{41},
\end{equation}
\begin{equation}
\dot{\rho_{32}}=-[\gamma_{32}+i(\Delta_{c}-\Delta_{p}+\delta)]\rho_{32}-i\Omega_{c}\rho_{34}-i\Omega_{p}\rho_{31}+i\Omega_{B}\rho_{42},
\end{equation}
\begin{equation}
\dot{\rho_{33}}=-\Gamma(\rho_{33}-\rho_{11})-\Gamma_{31}\rho_{33}-\Gamma_{32}\rho_{33}+\Gamma_{43}\rho_{44}+i\Omega_{B}(\rho_{43}-\rho_{34}),
\end{equation}
\begin{equation}
\dot{\rho_{41}}=-[\gamma_{41}+i(\Delta_{p}+\Delta_{c})]\rho_{41}+i\Omega_{c}\rho_{21}-i\Omega_{p}\rho_{42}+i\Omega_{B}\rho_{31},
\end{equation}
\begin{equation}
\dot{\rho_{42}}=-(\gamma_{42}+i\Delta_{c})\rho_{42}+i\Omega_{c}(\rho_{22}-\rho_{44})-i\Omega_{p}\rho_{41}+i\Omega_{B}\rho_{32},
\end{equation}
\begin{equation}
\dot{\rho_{43}}=-[\gamma_{43}+i(\Delta_{p}-\delta)]\rho_{43}+i\Omega_{c}\rho_{23}+i\Omega_{B}(\rho_{33}-\rho_{44}),
\end{equation}

where the above density matrix elements obey the
conditions:$\rho_{11}+\rho_{22}+\rho_{33}+\rho_{44}$=1 and
$\rho_{ij}$=$\rho_{ji}^{\ast}$.And $\Gamma_{ij}$(i,j=1,2,3,4)is the
spontaneous emission decay rate from level $|i\rangle$ to
level$|j\rangle$,ignoring the collision broaden
effect.$\gamma_{21}$=$\Gamma_{21}$/2,$\gamma_{31}$=($\Gamma_{31}+\Gamma_{32}$)/2,
$\gamma_{41}$=($\Gamma_{43}+\Gamma_{42}$)/2,
$\gamma_{42}$=($\Gamma_{43}+\Gamma_{31}+\Gamma_{21}$)/2,
$\gamma_{43}$=($\Gamma_{43}+\Gamma_{42}+\Gamma_{31}+\Gamma_{32}$)/2,$\gamma_{32}$=
($\Gamma_{32}+\Gamma_{31}+\Gamma_{21}$)/2 are the decay rates to the
corresponding transitions.The detuning of the fields defined as
$\Delta_{p}$=$\omega_{21}$-$\omega_{p}$,$\Delta_{c}$=$\omega_{42}$-$\omega_{c}$,
$\Delta_{B}$=$\omega_{43}$-$\omega_{P}$,respectively.$\omega_{ij}$=$\omega_{i}$-$\omega_{j}$
is the transition frequency of level$|i\rangle$ and $|j\rangle$ (i,
j=1,2,3,4) and we have $\delta$ =$\Delta_{p}$-$\Delta_{B}$.

According to the classical electromagnetic theory,the electric
polarizability is a rank 2 tensor defined by its Fourier transform
$\vec{P}_{e}(\omega_{P})$
$=\epsilon_{0}$$\alpha_{e}(\omega_{P})$$\vec{E}(\omega_{P})$,which
is calculated as the mean value of the atomic electric-dipole moment
operators by the definition $\vec{P}_{e}$
=Tr$\{$${\hat{\rho}\vec{d}}$$\}$=$\rho_{12}d_{21}$+c.c.where Tr
stands for trace.In the following, we only consider the
polarizability at the frequency $\omega_{P}$ of the incoming field
$\vec{E}_{p}$.Therefore we drop the explicit $\omega_{P}$ dependence
$\alpha_{e}(\omega_{P})\equiv\alpha_{e}$.Moreover,we choose
$\vec{E}_{p}$ parallel to the atomic dipole $\vec{d}_{21}$ so that
$\alpha_{e}$is a scalar,and its expression is as follows:

\begin{equation}
\alpha_{e}=\frac{\vec{d}_{21}\rho_{12}}{\epsilon_{0}\vec{E}_{p}}=\frac{\mid
{d_{21}}\mid^{2} \rho_{12}}{\epsilon_{0}\hbar\Omega_{p}},
\end{equation}

In the same way, the classical  magnetic polarizations of the medium
$\vec{P}_{m}(\omega_{P})$=$\mu_{0}\alpha_{m}\vec{E}(\omega_{P})$,which
is related to the mean value of the atomic dipole moment operator
through $\vec{P}_{m}$
=Tr$\{$${\hat{\rho}\vec{\mu}}$$\}$=$\rho_{34}$$\mu_{43}+c.c
$.According to the classical Maxwell's electromagnetic wavevector
relation,we choose magnetic dipole is perpendicular to the induced
electric dipole so that the magnetizability $\alpha_{m}$ is scalar,
and its expression is as follows:
\begin{equation}
\alpha_{m}=\frac{\mu_{0}\vec{\mu}_{43}\rho_{34}}{\vec{B}_{p}}=\frac{\mu_{0}\mid\mu_{43}\mid^{2}\rho_{34}}{\hbar\Omega_{B}}.
\end{equation}
According to the Clausius-Mossotti relations considering the local
effect in dense medium[25], the relative permittivity and relative
permeability are expressed as [18,26]
\begin{eqnarray}
\epsilon_{r}=\frac{1+\frac{2}{3}N\alpha_{e}}{1-\frac{1}{3}N\alpha_{e}},
\end{eqnarray}
\begin{eqnarray}
\mu_{r}=\frac{1+\frac{2}{3}N\gamma_{m}}{1-\frac{1}{3}N\gamma_{m}}.
\end{eqnarray}

In the above, we obtained the expressions for the electric
permittivity and magnetic permeability of the atomic media.In the
section that follows, we will demonstrate that the simultaneously
negative both permittivity and permeability,negative refraction
without absorption can be observed in the four-level atomic system.

\section{Results and discussion}

In the following,with the stationary solutions to the density-matrix
equations(2)-(10), we explore the property of simultaneously
negative both electric permittivity and magnetic permeability
through the numerical calculations.And several typical parameters
should be selected before the calculation.The parameters for the
electric and magnetic polarizabilities of atoms can be chosen as:
electric and magnetic transition dipole moments
$d_{21}$=2.5$\times$$10^{-29}$C$\cdot$m and $\mu_{34}$=7.0
$\times$$10^{-23}$C$\cdot m^{2}s^{-1}$[16], respectively.The density
of atoms N was choosed to be $6.5\times10^{25}m^{-3}$[27,28].And the
other parameters are scaled by $\gamma=10^{6}s^{-1}$:
$\Gamma_{31}$=$\Gamma_{41}$=$\Gamma_{42}$=$\Gamma_{43}$=
0.01$\gamma$,$\Gamma_{21}$=0.5$\gamma$, $\Gamma_{32}$=1$\gamma$. The
Rabi frequency of the probe field is $\Omega_{p}=0.5\gamma$ ,and
$\delta$ = - 20$\gamma$ without the condition of the two transition
frequencies $\omega_{43}$=$\omega_{21}$. The incoherent pumping rate
is $\Gamma=1.0\gamma$. The strong coherent optical field couples the
atomic system with variational Rabi frequencies
$\Omega_{c}=15\gamma$, 18$\gamma$,21$\gamma$, 24$\gamma$ and fixed
frequency detuning $\Delta_{c}$=20$\gamma$.

\begin{center}
\begin{figure}[h!]
     \centering
  \includegraphics[width=0.45\columnwidth]{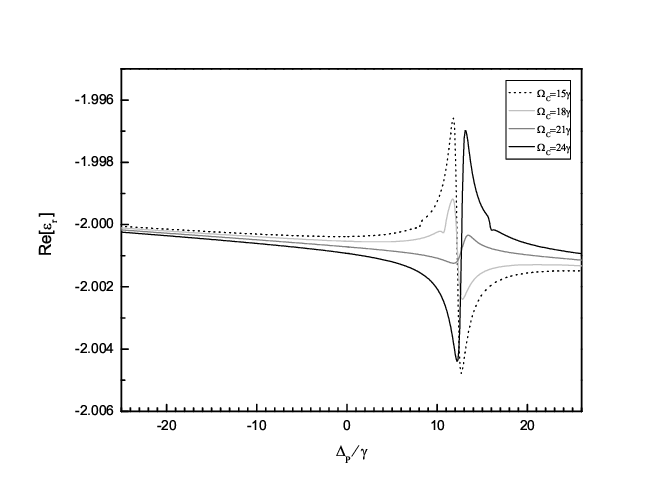}%
   \includegraphics[width=0.45\columnwidth]{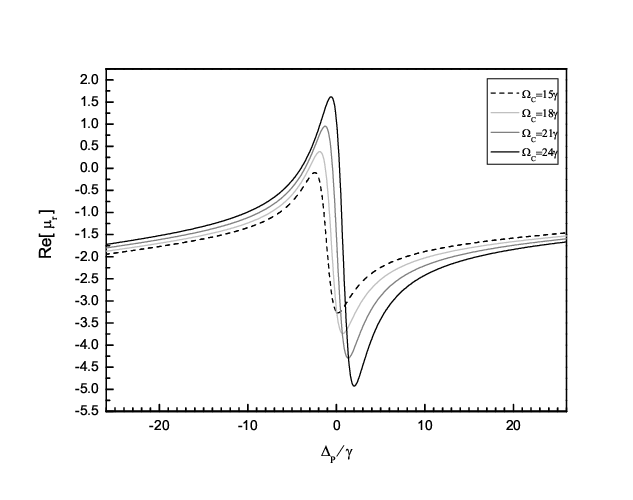}%
  \hspace{0in}%
  \includegraphics[width=0.45\columnwidth]{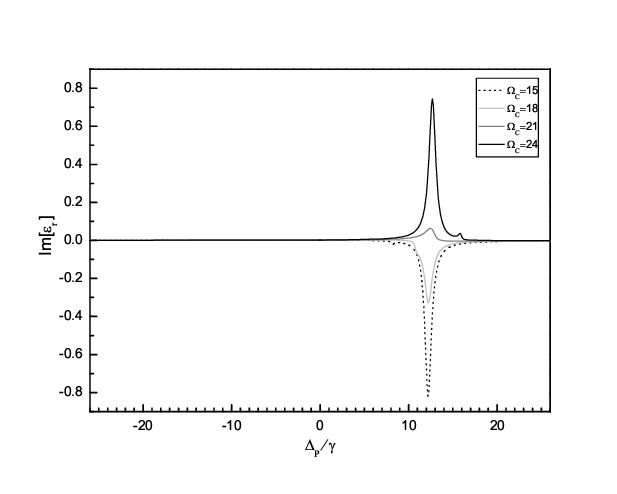}%
    \includegraphics[width=0.45\columnwidth]{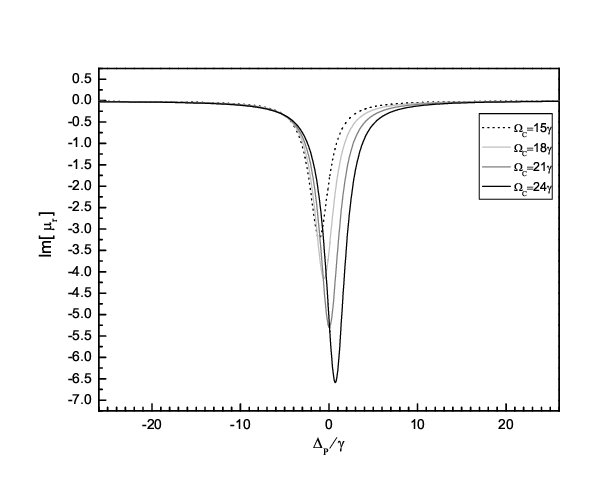}%
    \caption{The relative electric permittivity $\epsilon_{r}$ and magnetic
permeability $\mu_{r}$ of the system as a function of probe
frequency detuning $\Delta_{p}/\gamma$ for different values of
$\Omega_{C}$.}
\end{figure}\label{Fig.2}
\end{center}

Figure 2 shows the calculated electric permittivity $\epsilon_{r}$
and magnetic permeability $\mu_{r}$ as a function of probe field
detuning.Under the cooperation of the strong coherent field and
incoherent pumping,the dense four-level atomic system exhibits
left-handedness with simultaneously negative permittivity and
permeability at some frequency extents of the probe field.From the
profiles of the real part of the relative electric permittivity
$\epsilon_{r}$,we observe their values being proximately equal -2 in
the probe frequency range[-26$\gamma$,26$\gamma$],although the phase
of the figures are proximately opposite at $\Delta_{p}$=12.5$\gamma$
when the Rabi frequency $\Omega_{C}$ varies by 15$\gamma$,
18$\gamma$, 21$\gamma$ and 24$\gamma$.The influence of the strong
coherent field on the real part of the relative magnetic
permeability $\mu_{r}$ is different.When $\Omega_{C}$=15$\gamma$,the
value is all negative in the probe frequency extent.When
$\Omega_{C}$=18$\gamma$,21$\gamma$, 24$\gamma$,the positive values
emerge in the ranges of [-3$\gamma$, -1.1$\gamma$], [-3.4$\gamma$,
-0.4$\gamma$], [-3.85$\gamma$,0.36$\gamma$],respectively.And their
amplitude values are increasing with the variation of
$\Omega_{C}$.However,the ranges for simultaneously negative electric
permittivity $\epsilon_{r}$ and magnetic permeability $\mu_{r}$
still exist.In Figure 2,the transparency is shown by the imaginary
part of both the relative electric permittivity $\epsilon_{r}$ and
the relative magnetic permeability $\mu_{r}$in some frequency
extents.This is a very significant result for us.As is well known,
the photon absorption of atom can be greatly depressed via
electromagnetically induced transparency (EIT)[29].And the zero
absorption phenomena may occur in the EIT extents.With observation
of the profile of the imaginary part of $\epsilon_{r}$, the
excessive strong coherent field intensity causes the shrinking gain
and increasing absorption at $\Delta_{p}$=12.5$\gamma$.However,the
imaginary part of $\mu_{r}$ displays increasing gain,and its
amplitude is gradually amplified and gradually approaching to the
resonant point when the coherent field is varied the Rabi
frequencies by 15$\gamma$, 18$\gamma$,21$\gamma$ and 24$\gamma$.
Comparing the images of the relative electric permittivity
$\epsilon_{r}$ with the relative magnetic permeability $\mu_{r}$ in
Figure 2,we notice that the magnetic response of the probe field is
stronger than the electric response.The reason may come from that:
the magnetic component of the probe field couples levels $|4\rangle$
and $|3\rangle$,and the strong coherent field drives levels
$|2\rangle$ and $|4\rangle$.The effect of quantum coherence caused
by the strong coherent field on the magnetic component is stronger
than on the electric component which couples levels $|2\rangle$ and
$|1\rangle$.

\begin{center}
\begin{figure}[h!]
  \centering
  \includegraphics[width=0.45\columnwidth]{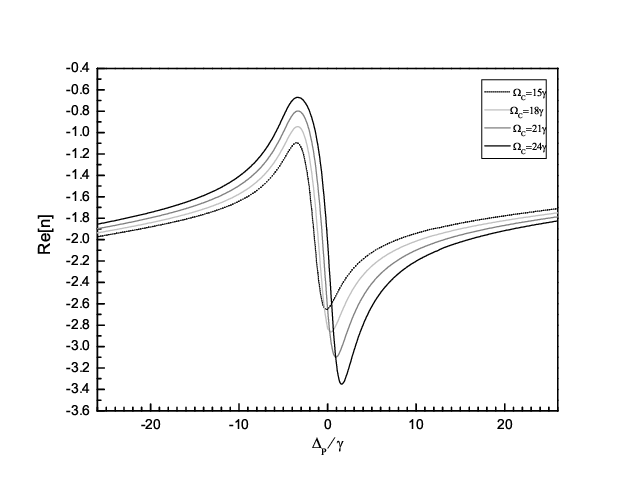}%
  \hspace{0in}%
  \includegraphics[width=0.45\columnwidth]{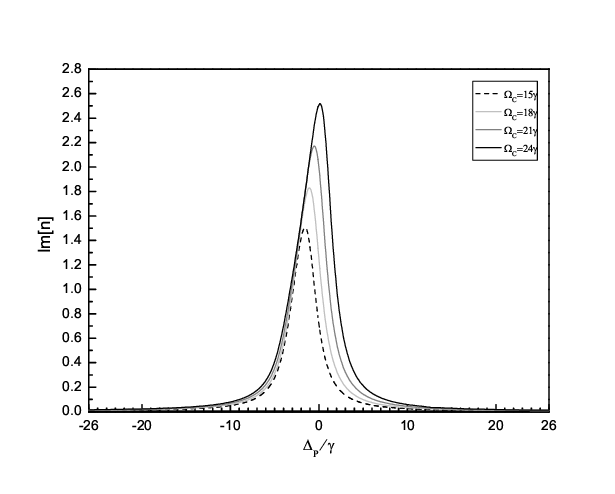}%
  \hspace{0in}%
  \caption{The refractive index as a function of the probe field
detuning $\Delta_{p}/\gamma$for different values of $\Omega_{C}$.}
\end{figure}\label{Fig.3}
\end{center}

\begin{center}
\begin{figure}[h!]
  \centering
  \includegraphics[width=0.45\columnwidth]{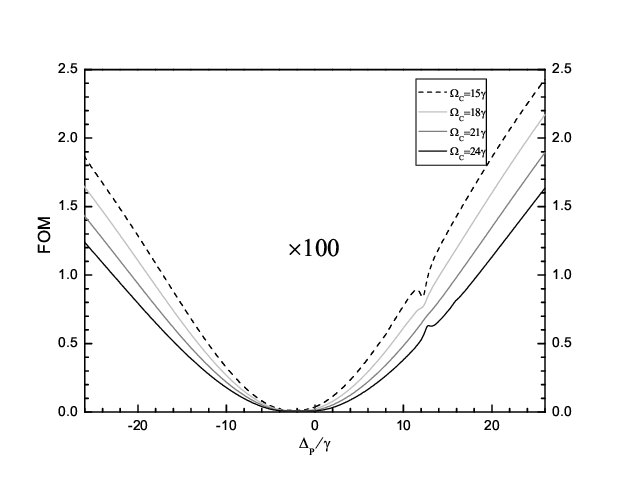}%
  \hspace{0in}%
    \caption{The figure of merit (FOM: $|real(n)/imag(n)|$) as a function of
the  detuning parameter detuning $\Delta_{p}/\gamma$for  different
values of $\Omega_{C}$.}
\end{figure}\label{Fig.4}
\end{center}

In Figure 3,the refraction index according to definition of the
left-handed material ( $n(\omega)=-\sqrt{\epsilon_{r}(\omega)
\mu_{r}(\omega)}$ ) [2] is plotted for different values of parameter
$\Omega_{C}$.As observed in Figure 3,the real part of the refractive
index shows negative values and their amplitudes enlarge gradually
with the increasing of $\Omega_{C}$. The imaginary part of the
refractive index displays absorption enhancing near the resonant
point and absorption depression on the both sides,when the coherent
field is varied the Rabi frequencies by 15$\gamma$,
18$\gamma$,21$\gamma$ and 24$\gamma$.The figure of
merit(FOM)$|Re(n)/Im(n)|$ shows how much the absorption is
suppressed .When the FOM is much larger than unity, it means that
there is almost no absorption in this area. As shown in Figure 4,
the FOM is far less than unity in the area near the resonant
point.This illustrates that the increasing absorption occurs in this
area ,which is consistent with the peaks of Im[n] near the resonance
in Figure 3. And the FOM has the largest value at the identical
re-scaled detuning parameter $\Delta_{p}/\gamma$ when $\Omega_{C}$=
15$\gamma$. It means that the absorption is depressed deeply when
the coherent field varies its Rabi frequency to $\Omega_{C}$=
15$\gamma$,and the excessive coherent field intensity wouldn't help
to realize zero absorption.The markedly feature of present scheme is
the absorption depression almost to zero in the left-handed atomic
system. One may get the reason for this from the results shown in
Figure 2:Because of the EIT effect, the electric and magnetic
non-absorption extents appear in the re-scaled detuning parameter
$\Delta_{p}/\gamma$ extent,which brings about the zero value of the
imaginary part of refraction index. As a result, the negative
refraction without absorption occurs in the left-handed atomic
system .

In experimental investigation, the solid sample of $E_{r}^{3+}$
doped into calcium fluorophosphate at room temperature might be a
good candidate because it has abundant energy levels, various
electric magnetic transitions and high density in the order
$10^{25}m^{-3}$[27,28].And the four levels $|1\rangle$,$|2\rangle$,
$|3\rangle$,and $|4\rangle$ in Figure 1 correspond to the energy
level configuration $I_{15/2}^{4}$,$I_{13/2}^{4}$, $I_{9/2}^{4}$ and
$I_{11/2}^{4}$ of the rare earth ion $E_{r}^{3+}$ doped in calcium
fluorophosphate, respectively.

\section{Conclusion}

In conclusion,we have demonstrated a scheme for realizing negative
refraction without absorption via an incoherent pump field and a
strong coherent field in the dense four-level atomic system.With the
application of the incoherent pump field to manipulate the
population distribution of each level and the variable strong
coherent field to create quantum coherence,the magnetic response is
amplified and the probe field propagates transparently in some
frequency extents.Without the constraint condition of two equal
transition frequencies responding to the probe field,the atomic
system displays negative refraction with vanishing absorption and
left-handedness. Therefore,our aim for searching the low-loss
negative refraction is possible by choosing appropriate parameters
in the scheme,in respect that the main applied limitation of the
negative refractive materials is the large amount of dissipation and
absorption.However,an excess coherent field intensity would increase
the absorption near the resonance.This can also be obviously
observed here.

\section*{Acknowledgments}

The work is supported by the National Natural Science Foundation of
China ( Grant No.60768001 and No.10464002 ).


\begin{thebibliography}{0}%
\makeatletter
\providecommand \@ifxundefined [1]{%
 \@ifx{#1\undefined}
}%
\providecommand \@ifnum [1]{%
 \ifnum #1\expandafter \@firstoftwo
 \else \expandafter \@secondoftwo
 \fi
}%
\providecommand \@ifx [1]{%
 \ifx #1\expandafter \@firstoftwo
 \else \expandafter \@secondoftwo
 \fi
}%
\providecommand \natexlab [1]{#1}%
\providecommand \enquote  [1]{``#1''}%
\providecommand \bibnamefont  [1]{#1}%
\providecommand \bibfnamefont [1]{#1}%
\providecommand \citenamefont [1]{#1}%
\providecommand \href@noop [0]{\@secondoftwo}%
\providecommand \href [0]{\begingroup \@sanitize@url \@href}%
\providecommand \@href[1]{\@@startlink{#1}\@@href}%
\providecommand \@@href[1]{\endgroup#1\@@endlink}%
\providecommand \@sanitize@url [0]{\catcode `\\12\catcode `\$12\catcode
  `\&12\catcode `\#12\catcode `\^12\catcode `\_12\catcode `\%12\relax}%
\providecommand \@@startlink[1]{}%
\providecommand \@@endlink[0]{}%
\providecommand \url  [0]{\begingroup\@sanitize@url \@url }%
\providecommand \@url [1]{\endgroup\@href {#1}{\urlprefix }}%
\providecommand \urlprefix  [0]{URL }%
\providecommand \Eprint [0]{\href }%
\providecommand \doibase [0]{http://dx.doi.org/}%
\providecommand \selectlanguage [0]{\@gobble}%
\providecommand \bibinfo  [0]{\@secondoftwo}%
\providecommand \bibfield  [0]{\@secondoftwo}%
\providecommand \translation [1]{[#1]}%
\providecommand \BibitemOpen [0]{}%
\providecommand \bibitemStop [0]{}%
\providecommand \bibitemNoStop [0]{.\EOS\space}%
\providecommand \EOS [0]{\spacefactor3000\relax}%
\providecommand \BibitemShut  [1]{\csname bibitem#1\endcsname}%
\let\auto@bib@innerbib\@empty
\end{thebibliography}%


\begin{thebibliography}{00}
\bibitem{1}M.O.Scully,M.S.Zubairy, {\it Quantum Optics(Cambridge University Press, Cambridge, 1997)}.
\bibitem{2}V.G.Veselago,{\it Sov.Phys.Usp.} {\bf 10} (1968) 509-514 .
\bibitem{3}J.B.Pendry,{\it Phys.Rev.Lett.} {\bf 85} (2000) 3966-3969.
\bibitem{4}L.Chen,S.He,L.Shen,{\it Phys.Rev.Lett.} {\bf 92} (2004)107404.
\bibitem{5}K.Aydin,I.Bulu,E.Ozbay,{\it Appl.Phys.Lett} {\bf 90} (2007) 254102
\bibitem{6}P.R.Berman, {\it Phys.Rev.E} {\bf 66} (2002) 067603
\bibitem{7}Y.P.Yang,J.P.Xu,H.Chen,and S.Y.Zhu,{\it Phys.Rev.Lett.} {\bf 100} (2008) 043601.
\bibitem{8}V.Yannopapas,E.Paspalakis,and N.V.Vitanov,{\it Phys.Rev.Lett.} {\bf 103} (2008) 063602.
\bibitem{9}Z.M.Zhang,C.J.Fu, {\it Appl. Phys. Lett.} {\bf 80} (2002) 1097-1099.
\bibitem{10}R.A.Shelby,D.R.Smith,S.Schultz,{\it Science} {\bf 292} (2001) 77-79.
\bibitem{11}J.Pendry,{\it Nature} {\bf 423} (2003) 22-23.
\bibitem{12}E.Cubukcu,{\it Nature} {\bf 423} (2003) 604-605.
\bibitem{13}G.V.Eleftheriades,A.K.Iyer,P.C.Kremer,{\it IEEE Trans.Microwave Theory Tech.} {\bf 50}(2002)2702-2712.
\bibitem{14}J.B.Pendry,{\it Science} {\bf 306} (2004) 1353-1355.
\bibitem{15}V.Yannopapas,{\it J. Phys.: Condens. Matter} {\bf 18}(2006) 6883-6890
\bibitem{16}Q.Thommen,P.Mandel,{\it Phys.Rev.Lett.} {\bf 96}(2006) 053601.
\bibitem{17}M.$\ddot{o}$.Oktel,$\ddot{o}$.E.M$\ddot{u}$tecapl$\check{g}$u,{\it Phys.Rev.A} {\bf 70}(2004)053806.
\bibitem{18}C.S.Zhao,D.Z.Liu, {\it Int.J.Quant.Inf.} {\bf 7} (2009) 747-754.
\bibitem{19}J.Q.Shen,{\it Phys.Lett.A} {\bf 357}(2006) 54-60.
\bibitem{20}J.K$\ddot{a}$stel,M.Fleischhauer,S.F.Yelin,R.L.Walsworth,{\it Phys.Rev.Lett.} {\bf 99}(2007) 073602.
\bibitem{21}J.K$\ddot{a}$stel,M.Fleischhauer,S.F.Yelin,R.L.Walsworth,{\it Phys.Rev. A} {\bf 79}(2009) 063818.
\bibitem{22}P.Tassin,L.Zhang,Th.Koschny,E.N.Economou,and C.M.Soukoulis,{\it Phys.Rev.Lett.} {\bf 102}(2009) 053901.
\bibitem{23}F.L. Li,A.P. Fang,M.Wang,{\it J. Phys.B}{\bf 42}(2009) 199505.
\bibitem{24}H. J. Zhang,Y.P Niu,S. Q. Gong,{\it Phys. Lett. A} {\bf 363}(2007) 497-501.
\bibitem{25}G. S. Agarwal,R.W.Boyd,{\it Phys.Rev.A } {\bf 60}(1999) R2681.
\bibitem{26}J. Q. Shen,{\it J.Mod.Opt.}{\bf 53} (2006) 2195-2205.
\bibitem{27}A. A. Kaminskii, V. Mironov, S. A. Kornienko, S. N. Bagaev,G. Boulon, A. Brenier,B. Di Bartolo, {\it Phys.Status Solidi A }{\bf 151}(1995) 231-255.
\bibitem{28}D.K.Sardar,C.H.Coeckelenbergh,R.M.Yow,J.B.Gruber,T.H.Allik,{\it J.Appl.Phys.}{\bf98}(2005) 033535.
\bibitem{29}N.Papasimakis,V.A.Fedotov,N.I.Zheludev,S.L.Prosvirnin{\it Phys.Rev.Lett.} {\bf 101}(2008) 253903.
\end{thebibliography}
\end{document}